\documentclass[article,12pt,superscriptaddress,showkeys,showpacs] {revtex4}
\usepackage{amsfonts}
\usepackage{graphicx, amsmath}
\begin{document}

\title[Short Title]{Direct conversion of a three-atom W state to a Greenberger-Horne-Zeilinger state in spatially separated cavities}

\author{Guo-Yuan Wang}
\affiliation{Department of
Physics, College of Science, Yanbian University, Yanji, Jilin
133002, People's Republic of China}
\author{Dong-Yang Wang}
\affiliation{Department of
Physics, College of Science, Yanbian University, Yanji, Jilin
133002, People's Republic of China}
\author{Wen-Xue Cui}
\affiliation{Department of
Physics, College of Science, Yanbian University, Yanji, Jilin
133002, People's Republic of China}
\author{Hong-Fu Wang\footnote{E-mail: hfwang@ybu.edu.cn}}
\affiliation{Department of Physics, College of Science, Yanbian University, Yanji, Jilin 133002, People's Republic of China}
\affiliation{School of Physics, Northeast Normal University, Changchun, Jilin 130024, People's Republic of China}
\author{Ai-Dong Zhu}
\affiliation{Department of
Physics, College of Science, Yanbian University, Yanji, Jilin
133002, People's Republic of China}
\author{Shou Zhang\footnote{E-mail: szhang@ybu.edu.cn}}
\affiliation{Department of
Physics, College of Science, Yanbian University, Yanji, Jilin
133002, People's Republic of China}

\begin{abstract}
State conversion between Greenberger-Horne-Zeilinger (GHZ) state and W state is an open challenging problem because they cannot be converted to each other only by local operations and classical communication. Here we propose a cavity quantum electrodynamics method based on interference of polarized photons emitted by the atoms trapped in spatially separated optical cavities that can convert a three-atom W state to a GHZ state. We calculate the success probability and fidelity of the converted GHZ state when the cavity decay, atomic spontaneous decay, and photon leakage of the cavities are taken into account for a practical system, which shows that the proposed scheme is feasible and within the reach of current experimental technology.

\end{abstract}

\maketitle
\section{Introduction}\label{sec0}
Quantum entanglement, which is recognized as an essential ingredient for testing local hidden variable theories against quantum mechanics, has extensive application in quantum computing and quantum information processing. It is well known that multipartite entangled states have many properties more peculiar than the bipartite ones because they exhibit the contradiction between local hidden variable theories and quantum mechanics even for nonstatistical predictions, as opposed to the statistical ones for the Einstein-Podolsky-Rosen (EPR) states~\cite{DPTAJP904358, AFPRA0469}. It has been shown that genuine three-qubit entanglement comes in two different inconvertible types represented by the GHZ state and the W state~\cite{DMAAAJP9058, WGJPRA0062}. The GHZ state is inequivalent to the W state in the sense that they cannot be converted into each other even under any stochastic local operations and classical communication (SLOCC). These two kinds of entangled states represent two distinct classes of three-qubit entanglement and can perform different quantum information processing tasks, and much interest has been paid in the investigation of how to convert the two types of entanglement into each other. Based on positive operator valued measures (POVMs), Walther {\it et al.} proposed a method to convert a GHZ state to an arbitrarily good approximation to a W state and experimentally realized this scheme in the three-photon case~\cite{PKAPRL0594}. In Ref.~\cite{TTSTMNPRL09102}, the authors experimentally demonstrated a transformation of two Einstein-Podolsky-Rosen photon pairs distributed among three parties into a three-photon W state using local operations and classical communication. We also proposed a linear optical method to convert $N-1$ $(N\geq3)$ entangled two-photon pairs distributed among $N$ parties into a $N$-photon W state~\cite{HIJTP1352}. Through a dissipative dynamics process in an open quantum system, Song {\it et al.} showed that a four-atom W state can be converted into a GHZ state with deterministic probability~\cite{JXQLYHPRA1388}. Furthermore, we proposed a linear-optics-based scheme for local conversion of four Einstein-Podolsky-Rosen photon pairs distributed among five parties into four-photon polarization-entangled decoherence-free states using SLOCC and non-photon-number-resolving detectors~\cite{HSAXKOE1119}. These works make it possible to convert different kinds of quantum states into each other.

Cavity quantum electrodynamics (QED) system is a promising candidate for quantum information processing because atoms are suitable for storing information in stationary nodes and photons suitable for transporting information. In practice, however, the distribution of entanglement for atoms over long distance is difficult because of unavoidable transmission losses and decoherence in the quantum channel. In recent years, several schemes have been proposed to realize quantum computation and engineer entanglement between atoms trapped in distant optical cavities, either through detection of leaking photons~\cite{HXYSKJPB0942, SPMVPRL9983, LHPRL0390, DMSPRL0391, HXKJPB0942}, or through direct connection of the cavities by optical fiber~\cite{TPRL9779, JPHHPRL9778, ASSPRL0696, ZFPRA0775, JYHEPL0987, SAPL0994, SCFPRA1082, SCPB1019, HSAKJOSAB1229}. In this paper, we propose a scheme for converting a three-atom W state to a GHZ state with the certain success probability. In the scheme each $\lambda\lambda$-type atom is individually trapped in an optical cavity and by the interference and detections of the polarized photons leaking out of the separate optical cavities, the conversion from a three-atom W state to a GHZ state is achieved. The scheme does not require the simultaneous click of the detectors and it is robust against the asynchronous emission of the polarized photons and the detection inefficiency. Furthermore, we consider the influence of cavity decay, atomic spontaneous decay, and photon leakage on the success probability of the scheme and the fidelity of the state for a practical system. Compared with Ref.~\cite{PKAPRL0594}, our scheme has the following merits: (i) the method proposed in Ref.~\cite{PKAPRL0594} is to convert a three-photon GHZ state to a approximate W state, while in our scheme, a three-atom W state can be converted to a exact GHZ state; (ii) in the ideal case, the fidelity of the converted state in Ref.~\cite{PKAPRL0594} is 3/4, while in our scheme, the fidelity is 1.0.

\begin{figure}
\includegraphics[width=5.3in]{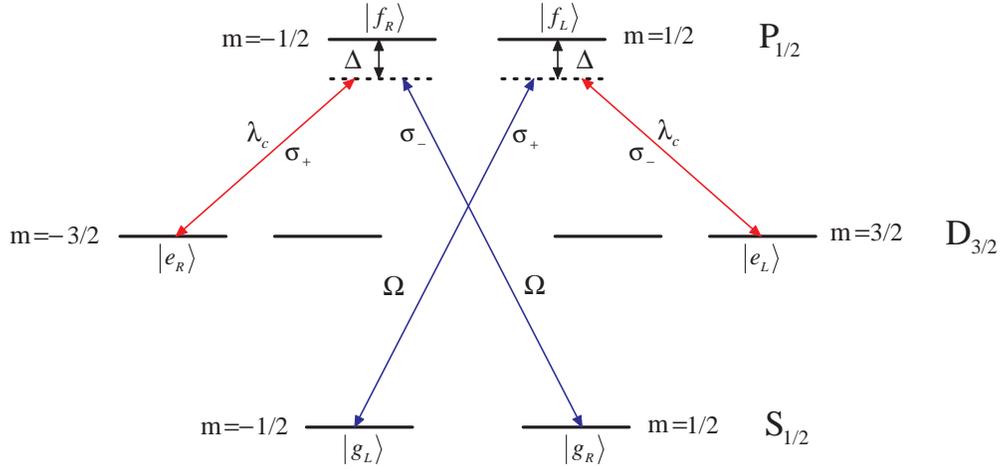}\caption{(Color online) The level configuration and excitation scheme of the atoms. $|g_{L}\rangle,~|g_{R}\rangle,~|e_{L}\rangle$, and $|e_{R}\rangle$ are the ground states, $|f_{L}\rangle$ and $|f_{R}\rangle$ are the excited states. The transition $|g_L\rangle\leftrightarrow|f_L\rangle$ ($|g_R\rangle\leftrightarrow|f_R\rangle$) is driven by a classical field with left-circular (right-circular) polarization and the transition $|e_L\rangle\leftrightarrow|f_L\rangle$ ($|e_R\rangle\leftrightarrow|f_R\rangle$) is coupled to cavity mode $a_L$ ($a_R$) with the left-circular (right-circular) polarization.}
\end{figure}

\section{Conversion of a three-atom W state to a GHZ state by interference of polarized photons}\label{sec1}
We consider a $\lambda\lambda$-type atom, as shown in Fig.~1. This kind of level structure has been proposed to generate entangled single-photon wave packets~\cite{CKPJPPRA0061} and achieve quantum computation in a single
cavity~\cite{TSJPPRL9575, XWPRA0571}. The transition $|g_L\rangle\leftrightarrow|f_L\rangle$
($|g_R\rangle\leftrightarrow|f_R\rangle$) is driven by a classical field with left-circular (right-circular)
polarization. The transition $|e_L\rangle\leftrightarrow|f_L\rangle$ ($|e_R\rangle\leftrightarrow|f_R\rangle$)
is coupled to cavity mode $a_L$ ($a_R$) with the left-circular (right-circular) polarization. The Hamiltonian
of the system is written as
\begin{eqnarray}\label{e01}
H_{\rm I}=\sum\limits_{j=L,R}{\big[}\Delta|f_j\rangle\langle f_j|+(\lambda_ca_j|f_j\rangle\langle e_{j}|+\Omega|f_j\rangle\langle g_j|+{\rm H.c.}){\big]},
\end{eqnarray}
where $\Delta$ is the detuning between the cavity mode and the corresponding atomic transition, $\lambda_c$ is the coupling strength between the atom and cavity mode, $\Omega$ is the Rabi frequency of the classical field, and $a_L$ and $a_R$ are the annihilation operators of the left-circular and right-circular polarization
modes $L$ and $R$, respectively. Under the large-detuning conditions $\Delta\gg \lambda_c, \Omega$,
the excited states of the atom $|f_L\rangle$ and $|f_R\rangle$ are only virtually excited during the atom-cavity
interaction process. Therefore, the effect of rapidly oscillating terms can be neglected and the levels $|f_L\rangle$ and $|f_R\rangle$
can be eliminated adiabatically, leading to the effective Hamiltonian
\begin{eqnarray}\label{e02}
H_{\rm eff}=\sum\limits_{j=L,R}-\left[\frac{\lambda_c^2}{\Delta}|e_{j}\rangle\langle e_{j}|a_{j}^\dag a_{j}+\frac{\Omega^2}{\Delta}|g_j\rangle\langle g_j|+\frac{\lambda_c\Omega}{\Delta}{\big(}|g_j\rangle\langle e_{j}|a_{j}+{\rm H.c.}{\big)}\right].
\end{eqnarray}

\begin{figure}
\includegraphics[width=4.2in]{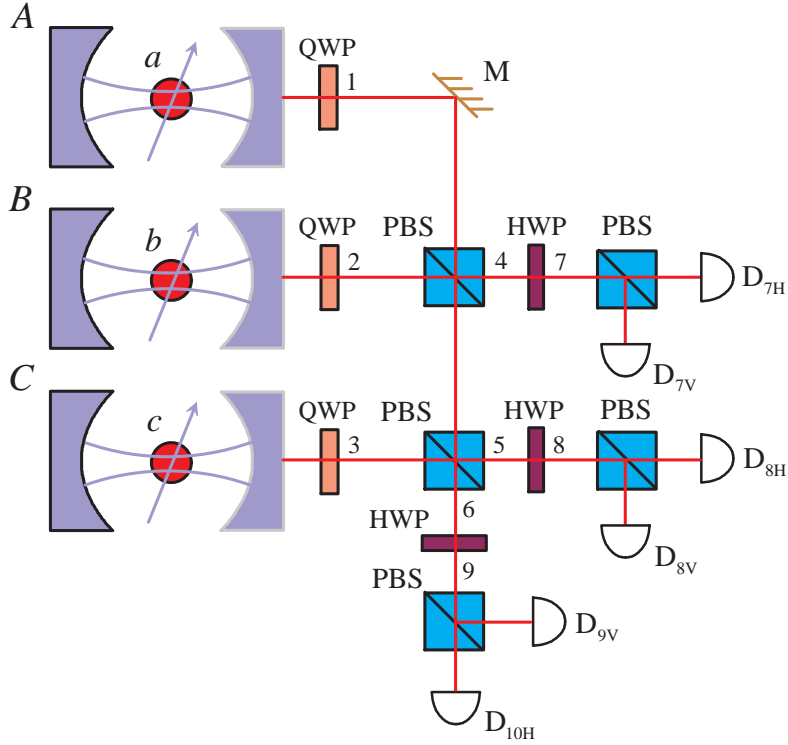}\caption{(Color online) Schematic setup of converting a three-atom W state to a GHZ state. Atoms $a$, $b$, and $c$ are trapped in three Fabry-P\'{e}rot cavities $A$, $B$, and $C$, respectively. Here QWP denotes a quarter-wave plate, PBS denotes a polarization beam splitter that transmits $H$ photon and reflects $V$ photon, HWP denotes a half-wave plate, $M$ denotes mirror, and $D$ is a conventional photon detector.}
\end{figure}

Assume that atoms $a$, $b$, and $c$, which are respectively trapped in three spatially
separated optical cavities $A$, $B$, and $C$, as shown in Fig.~2, are in the following three-atom entangled W state,
\begin{eqnarray}\label{e03}
|\Psi\rangle_{{\rm W}}=\frac{1}{\sqrt{3}}(|g_{L}\rangle_a|g_{L}\rangle_b|g_{R}\rangle_c+|g_{L}\rangle_a|g_{R}\rangle_b|g_{L}\rangle_c+|g_{R}\rangle_a|g_{L}\rangle_b|g_{L}\rangle_c).
\end{eqnarray}
This kind of W state can be prepared by using the same method proposed in Ref.~\cite{HXYSKJPB0942}. Performing a Hadamard gate operation, which can be achieved by a $\pi/2$ microwave pulse, on atoms $a$, $b$, and $c$ respectively, to accomplish the transformation
\begin{eqnarray}\label{e04}
|g_{L}\rangle&\rightarrow&\frac{1}{\sqrt{2}}(|g_{L}\rangle+|g_{R}\rangle),\cr\cr
|g_{R}\rangle&\rightarrow&\frac{1}{\sqrt{2}}(|g_{L}\rangle-|g_{R}\rangle).
\end{eqnarray}
After that, the state becomes
\begin{eqnarray}\label{e05}
|\Psi\rangle_1&=&\frac{1}{2\sqrt{6}}(3|g_{L}\rangle_a|g_{L}\rangle_b|g_{L}\rangle_c-3|g_{R}\rangle_a|g_{R}\rangle_b|g_{R}\rangle_c+|g_{L}\rangle_a|g_{L}\rangle_b|g_{R}\rangle_c
+|g_{L}\rangle_a|g_{R}\rangle_b|g_{L}\rangle_c\cr\cr&&-|g_{L}\rangle_a|g_{R}\rangle_b|g_{R}\rangle_c
+|g_{R}\rangle_a|g_{L}\rangle_b|g_{L}\rangle_c-|g_{R}\rangle_a|g_{L}\rangle_b|g_{R}\rangle_c-|g_{R}\rangle_a|g_{R}\rangle_b|g_{L}\rangle_c).
\end{eqnarray}
If optical cavities $A$, $B$, and $C$ are initially prepared in vacuum states $|0_L,0_R\rangle_A\otimes|0_L,0_R\rangle_B\otimes|0_L,0_R\rangle_C$,
after time $t$, the temporal evolution of the total system is expressed as
\begin{eqnarray}\label{e06}
|\Psi(t)\rangle_2&=&\frac{1}{2\sqrt{6}}{\big(}3|\phi_{L}(t)\rangle_a|\phi_{L}(t)\rangle_b|\phi_{L}(t)\rangle_c-3|\phi_{R}(t)\rangle_a|\phi_{R}(t)\rangle_b|\phi_{R}(t)\rangle_c
\cr\cr&&+|\phi_{L}(t)\rangle_a|\phi_{L}(t)\rangle_b|\phi_{R}(t)\rangle_c+|\phi_{L}(t)\rangle_a|\phi_{R}(t)\rangle_b|\phi_{L}(t)\rangle_c
\cr\cr&&-|\phi_{L}(t)\rangle_a|\phi_{R}(t)\rangle_b|\phi_{R}(t)\rangle_c+|\phi_{R}(t)\rangle_a|\phi_{L}(t)\rangle_b|\phi_{L}(t)\rangle_c
\cr\cr&&-|\phi_{R}(t)\rangle_a|\phi_{L}(t)\rangle_b|\phi_{R}(t)\rangle_c-|\phi_{R}(t)\rangle_a|\phi_{R}(t)\rangle_b|\phi_{L}(t)\rangle_c{\big)},
\end{eqnarray}
where
\begin{eqnarray}\label{e07}
|\phi_{L}(t)\rangle_\mu=\alpha|g_L\rangle_\mu|0_L,0_R\rangle_\nu+\beta|e_L\rangle_\mu|1_L,0_R\rangle_\nu,\cr\cr
|\phi_{R}(t)\rangle_\mu=\alpha|g_R\rangle_\mu|0_L,0_R\rangle_\nu+\beta|e_R\rangle_\mu|0_L,1_R\rangle_\nu,
\end{eqnarray}
with $\mu=a,b,c$, $\nu=A,B,C$, and
\begin{eqnarray}\label{e08}
\alpha&=&\frac{\lambda_c^2+\Omega^2\cos[(\lambda_c^2+\Omega^2)t/\Delta]+{\rm i}\Omega^2\sin[(\lambda_c^2
+\Omega^2)t/\Delta]}{\lambda_c^2+\Omega^2},\cr\cr
\beta&=&\frac{-\lambda_c\Omega+\Omega^2\cos[(\lambda_c^2+\Omega^2)t/\Delta]+{\rm i}\lambda_c\Omega\sin[(\lambda_c^2
+\Omega^2)t/\Delta]}{\lambda_c^2+\Omega^2}.
\end{eqnarray}
With the choices of $\lambda_c=\Omega$ and $t=\frac{\Delta\pi}{\lambda_c^2+\Omega^2}$,
then the photons leaking out from the cavities $A$, $B$, and $C$
first pass through a quarter-wave plate (QWP), whose action is to
make the left-circularly polarized photons become vertically
polarized photons and to make the right-circularly polarized photons become
horizontally polarized photons, i.e., $|1_L,0_R\rangle\rightarrow|V\rangle$
and $|0_L,1_R\rangle\rightarrow|H\rangle$,
respectively, as shown in Fig.~2. Next the photons in modes 1, 2, and 3 pass through
a series of polarization beam splitters (PBSs) and half-wave plates (HWPs). Here the
action of the PBS is to transmit the horizontal polarization and reflect vertical polarization and the action of the HWP is given by the transformation
\begin{eqnarray}\label{e9}
|H\rangle\rightarrow\frac{1}{\sqrt{2}}(|H\rangle+|V\rangle),\cr\cr
|V\rangle\rightarrow\frac{1}{\sqrt{2}}(|H\rangle-|V\rangle).
\end{eqnarray}
After that, the resulting state of the atom-photon system is given by
\begin{eqnarray}\label{e10}
|\Psi\rangle_r&=&\frac{1}{2\sqrt{6}}(-3|e_L\rangle_a|e_L\rangle_b|e_L\rangle_c|\psi_7^-\rangle|\psi_8^-\rangle|\psi_9^-\rangle
+3|e_R\rangle_a|e_R\rangle_b|e_R\rangle_c|\psi_7^+\rangle|\psi_8^+\rangle|\psi_9^+\rangle
\cr\cr&&-|e_L\rangle_a|e_L\rangle_b|e_R\rangle_c|\psi_7^-\rangle|\psi_8^-\rangle|\psi_8^+\rangle
-|e_L\rangle_a|e_R\rangle_b|e_L\rangle_c|\psi_7^-\rangle|\psi_7^+\rangle|\psi_9^-\rangle
\cr\cr&&+|e_L\rangle_a|e_R\rangle_b|e_R\rangle_c|\psi_7^-\rangle|\psi_7^+\rangle|\psi_8^+\rangle
-|e_R\rangle_a|e_L\rangle_b|e_L\rangle_c|\psi_8^-\rangle|\psi_9^-\rangle|\psi_9^+\rangle
\cr\cr&&+|e_R\rangle_a|e_L\rangle_b|e_R\rangle_c|\psi_8^-\rangle|\psi_8^+\rangle|\psi_9^+\rangle
+|e_R\rangle_a|e_R\rangle_b|e_L\rangle_c|\psi_7^+\rangle|\psi_9^-\rangle|\psi_9^+\rangle,
\end{eqnarray}
where $|\psi_m^\pm\rangle=1/\sqrt{2}(|H_m\rangle\pm|V_m\rangle)$, with $m\in\{7,8,9\}$.

Finally, the photons are detected by conventional photon detectors, which we consider here are realistic detectors commonly used
in photonic experiments. This kind of detector cannot
resolve the number of the detected photons but instead tell us
whether photons exist in a detection event with nonunit probability
$\eta_d$. Usually, the dark count of the detector is considerable low
and hence can be neglected. The positive-operator-valued-measure
(POVM) describing a conventional photon detector
is given by~\cite{HSPRA0979, HSAXKOE1119}
\begin{eqnarray}\label{e11}
\Pi_{{\rm
off}}&=&\sum\limits_{k=0}^\infty{\big(}1-\eta_d{\big)}^k|k\rangle\langle
k|,\cr\Pi_{{\rm
click}}&=&\sum\limits_{k=0}^\infty{\big[}1-{\big(}1-\eta_d{\big)}^k{\big]}|k\rangle\langle
k|.
\end{eqnarray}
Here $\Pi_{{\rm off}}$ is the POVM element for no photocounts and $\Pi_{{\rm click}}$ is that for photocounts. We only consider the event that one of the detectors ($D_{7H}$, $D_{7V}$) detects photons and another does not register any photon, similar events to detectors ($D_{8H}$, $D_{8V}$) and ($D_{9H}$, $D_{9V}$).
After the detection of photons, the state of atoms $a$, $b$, and $c$ is given by
\begin{eqnarray}\label{e12}
\rho^k_{{\rm out}}&=&\frac{{\rm
Tr}_{7H,7V,8H,8V,9H,9V}\left[\Pi_{{\rm
click}}^{7\delta}\Pi_{{\rm click}}^{8\delta}\Pi_{{\rm
click}}^{9\delta}\Pi_{{\rm
off}}^{7\gamma}\Pi_{{\rm off}}^{8\gamma}\Pi_{{\rm
off}}^{9\gamma}{\big(}|\Psi\rangle_r{_r\langle\Psi|}{\big)}\right]}{\
{\rm
Tr}_{a,b,c,7H,7V,8H,8V,9H,9V}\left[\Pi_{{\rm
click}}^{7\delta}\Pi_{{\rm click}}^{8\delta}\Pi_{{\rm
click}}^{9\delta}\Pi_{{\rm
off}}^{7\gamma}\Pi_{{\rm off}}^{8\gamma}\Pi_{{\rm
off}}^{9\gamma}{\big(}|\Psi\rangle_r{_r\langle\Psi|}{\big)}\right]\
}\cr\cr&=&|\Psi\rangle_a^j{_a^j\langle\Psi|},
\end{eqnarray}
where $\delta\neq\gamma\in\{H,V\}$, $j=1,2$, and
$|\Psi\rangle_r$ is denoted by Eq.~({\ref{e10}}), with
\begin{eqnarray}\label{e13}
|\Psi\rangle_a^1&=&\frac{1}{\sqrt{2}}(|e_L\rangle_a|e_L\rangle_b|e_L\rangle_c+|e_R\rangle_a|e_R\rangle_b|e_R\rangle_c),\cr\cr
|\Psi\rangle_a^2&=&\frac{1}{\sqrt{2}}(-|e_L\rangle_a|e_L\rangle_b|e_L\rangle_c+|e_R\rangle_a|e_R\rangle_b|e_R\rangle_c),
\end{eqnarray}
which are the three-atom GHZ states. Here $|\Psi\rangle_a^1$ corresponds to that photon detectors $\{D_{7H}, D_{8H}, D_{9V}\}$ (or $\{D_{7H}, D_{8V}, D_{9H}\}$, or $\{D_{7V}, D_{8H}, D_{9H}\}$, or $\{D_{7V}, D_{8V}, D_{9V}\}$) detect photons and the others do not register any photon, and $|\Psi\rangle_a^2$ corresponds to that photon detectors $\{D_{7H}, D_{8H}, D_{9H}\}$ (or $\{D_{7H}, D_{8V}, D_{9V}\}$, or $\{D_{7V}, D_{8H}, D_{9V}\}$, or $\{D_{7V}, D_{8V}, D_{9H}\}$) detect photons. The state $|\Psi\rangle_a^2$ can be transformed into the state $|\Psi\rangle_a^1$ by applying a classical microwave pulse to change the sign of an arbitrary atomic state.
The overall success probability for obtaining the state in Eq.~({\ref{e13}}) is
\begin{eqnarray}\label{e14}
P=\frac{3\eta_d^3}{4},
\end{eqnarray}
with $\eta_d$ being the quantum efficiency of photon detector. Performing the following transformations
\begin{eqnarray}\label{e15}
|e_L\rangle_i&\rightarrow&|g_L\rangle_i,\cr\cr
|e_R\rangle_i&\rightarrow&|g_R\rangle_i,~~i=a,b,c,
\end{eqnarray}
which can be achieved by applying fast Raman transitions to
manipulating atoms $a$, $b$, and $c$~\cite{HXYSKJPB0942, HASKJPB1043} individually.  Then the
state in Eq.~({\ref{e13}}) is mapped to the state
\begin{eqnarray}\label{e16}
|\Psi\rangle_{\rm GHZ}=\frac{1}{\sqrt{2}}(|g_L\rangle_a|g_L\rangle_b|g_L\rangle_c+|g_R\rangle_a|g_R\rangle_b|g_R\rangle_c).
\end{eqnarray}
In this way the conversion from a three-atom W state to a GHZ state is achieved, with the maximal success probability of being $3/4$ in an ideal case.

\section{The effects of cavity decay, spontaneous decay, and photon leakage}\label{sec2}
In this section, we investigate the influence of cavity decay, atomic spontaneous decay and photon leakage of the cavities. When the cavity decay is considered, the Hamiltonian is rewritten as
\begin{eqnarray}\label{e17}
H_{\rm eff}^\prime=H_{\rm eff}-{\rm i}\kappa\sum\limits_{j=L,R}a_j^\dag a_j,
\end{eqnarray}
where $H_{\rm eff}$ is denoted by Eq.~({\ref{e02}}) and  both the polarization modes $L$ and $R$ have been assumed to have the same loss rate $\kappa$. The evolution coefficients $\alpha$ and $\beta$ of the state denoted by Eq.~({\ref{e08}}) is thus given by
\begin{eqnarray}\label{e18}
\alpha^\prime&=&\frac{[\phi\cosh(\phi t/2)+\kappa\sinh(\phi t/2)]e^{\varphi t}}{\phi},\cr\cr
\beta^\prime&=&\frac{{\rm i}\eta(e^{\phi t}-1)e^{t(\varphi-\phi/2)}}{\phi},
\end{eqnarray}
where $\phi=\sqrt{\kappa^2-4\eta^2}$, $\eta=\frac{\lambda_c^2}{\Delta}$, $\varphi={\rm i}\eta-\frac{\kappa}{2}$,
and $\lambda_c=\Omega$. In this case the success probability corresponding to successful detections of photons
for obtaining the state $|\Psi\rangle_{\rm GHZ}$ in Eq.~({\ref{e16}}) at time $t$ is given by
\begin{eqnarray}\label{e19}
P_d=\frac{6\eta^6\left[1-\cos(\phi^\prime t)\right]^3e^{-3\kappa t}}{{\phi^\prime}^6},
\end{eqnarray}
where $\phi^\prime=\sqrt{4\eta^2-\kappa^2}$. The success probability $P_d$ as a function
of $\kappa t$ with different values $\eta$ was plotted in Fig.~3, which shows that the cavity decay rate $\kappa$ is the dominant noise source in the state conversion. The success probability $P_d$ (when $\kappa=0$, the maximal success probability is $P_d=0.75$) rapidly decreases to 0.03853 from 0.715584 when $\kappa t$ increases from 0.0156582 to 0.9896 (here we set $\eta=100\kappa$). Therefore, the waiting time for the effective detection of photons can be chosen to be a few times of cavity lifetime $1/\kappa$.
\begin{figure}
  \includegraphics[width=4.5in]{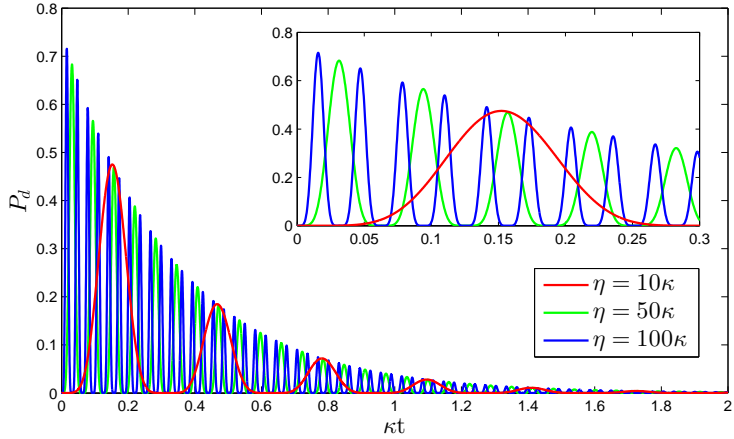}\\
  \caption{(Color online) The success probability $P_d$ as a function of $\kappa t$ with different values of $\eta$ for detecting three photons successfully in different modes shown in Fig.~2.}\label{f03}
\end{figure}

On the other hand, for convenience, we just consider the partial system including one atom and one cavity. Based on the density-matrix formalism, the master equation for the density matrices of the partial system can be expressed as
\begin{eqnarray}\label{e21}
\dot{\rho}&=&-{\rm i}[H_{\mathrm{I}},\rho]-\sum_{j=L, R}\Bigg[\frac{\kappa_{j}}{2}\left(a_{j}^{\dag}a_{j}\rho-2a_{j}\rho a_{j}^{\dag}+\rho a_{j}^{\dag}a_{j}\right)\cr\cr
&&+\sum_{x=g,e}\frac{\gamma_{j}^{fx}}{2}\left(\sigma_{j}^{ff}\rho-2\sigma_{j}^{xf}\rho\sigma_{j}^{fx}+\rho\sigma_{j}^{ff}\right)\Bigg],
\end{eqnarray}
where $H$ is denoted by Eq.~(\ref{e01}), $\kappa_{j}(j=L, R)$ denotes the decay rates of the cavity field, $\gamma_{j}^{fx}(x=g, e)$ denotes the spontaneous decay rate of the atom from level $|f_{j}\rangle$ to $|x_{j}\rangle$, $\sigma_{j}^{mn}=|m_{j}\rangle\langle n_{j}|(m,n=f,g,e)$ are the usual Pauli matrices. For the sake of convenience, we assume that $\gamma_{j}^{fx}=\gamma_{a}/2$ due to the equiprobably transition of $|f_{j}\rangle\leftrightarrow|x_{j}\rangle$ and $\kappa_{j}=\gamma_{j}^{fx}$ for simplicity. In the following, we analyze and discuss the parameter conditions and the experimental feasibility of the present scheme. With the choice of a scaling $\gamma$, then all parameters can be reduced to the dimensionless units related to $\gamma$. Setting $\Omega=2.9\gamma$, $\Delta=14\gamma$, and $\lambda_{c}=2.86\gamma$. By solving Eq.~(\ref{e21}) numerically, we obtain the effects of the atomic spontaneous decay and photon leakage of the cavities on the fidelity including three atoms and three cavities, as shown in Fig.~\ref{f03}.
In current experiments, the parameters $\lambda_{c}=2.5~\mathrm{GHz}, \kappa=10~\mathrm{MHz}$, and $\gamma_{a}=10~\mathrm{MHz}$ have been reported in Refs.~\cite{STKKEHPRA0571,MFMNP062}. For such parameters, the calculated fidelity is about $91.04\%$, which is relatively high. Even if the atomic spontaneous decay increases to $\lambda_{c}/\gamma_{a}=50$, the fidelity also can reach $90.09\%$.
\begin{figure}
  \includegraphics[width=5in]{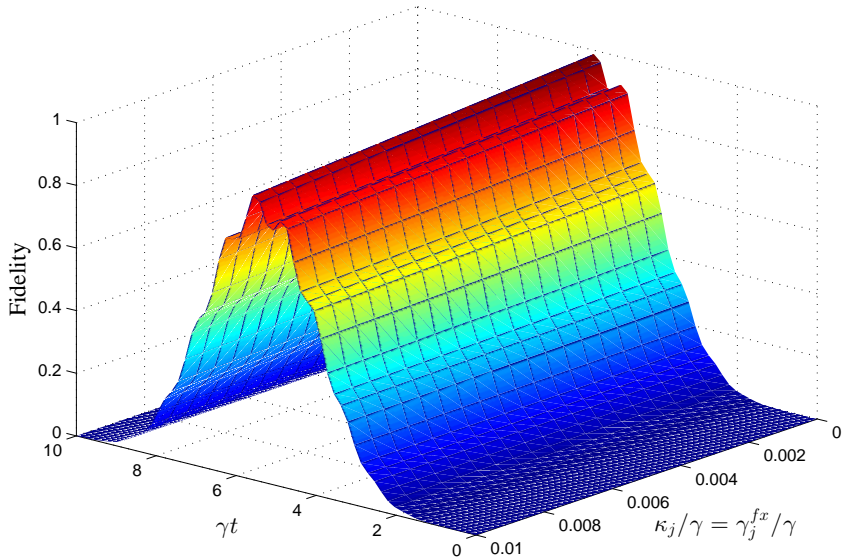}\\
  \caption{(Color online) The fidelity relates to the effects of the atomic spontaneous decay and the photon leakage of the cavities. The parameters are chosen as $\Omega=2.9\gamma$, $\Delta=14\gamma$, and $\lambda_{c}=2.86\gamma$.}\label{f03}
\end{figure}

\section{Conclusions}\label{sec3}
In conclusion, basing on the atom-cavity interaction and linear optical elements, we have proposed a method to convert a three-atom W state to a GHZ state by interference of polarized photons emitted by the atoms trapped in spatially separated cavities. In our scheme, the levels $|F=1/2,m=-1/2\rangle$ and $|F=1/2,m=1/2\rangle$ of
$4^2P_{1/2}$ for a $^{40}{\rm Ca}^+$ can be used as the excited states $|f_L\rangle$ and $|f_R\rangle$, $|F=1/2,m=1/2\rangle$ and $|F=1/2,m=-1/2\rangle$ of
$4^2S_{1/2}$ can be used as the ground states $|g_L\rangle$ and $|g_R\rangle$, and $|F=3/2,m=-3/2\rangle$ and
$|F=3/2,m=3/2\rangle$ of $3^2D_{3/2}$ can be used to serve as the states $|e_L\rangle$ and $|e_R\rangle$, respectively. The lifetimes
of the atomic levels $|e_L\rangle$, $|e_R\rangle$, $|g_L\rangle$, and $|g_R\rangle$ are comparatively long so that we can neglect the spontaneous decay
of these states. We analyze and discuss the effect of cavity decay on the success probability and the effects of spontaneous decay and photon leakage on the state fidelity, the calculated results show that our scheme might be experimentally realizable based on the current cavity QED and linear optical
techniques.

\begin{center}
{\small {\bf ACKNOWLEDGMENTS}}
\end{center}

This work was supported by the National Natural Science Foundation of China under
Grant Nos. 11264042, 11465020, 61465013, 11165015, and 11564041.

\end{document}